\documentclass{iopart}
\usepackage{graphicx}
\usepackage{epstopdf}

\usepackage{iopams}
\usepackage{color}

\begin{document}

\title[Multiple phase transitions in a TASEP with limited particles and fuel carriers]{Multiple phase transitions in a system of exclusion processes with limited reservoirs of particles and fuel carriers}
\author{Chris A Brackley$^1$, Luca Ciandrini$^1$ and M Carmen Romano$^{1,2}$}
\address{$^1$ Institute for Complex Systems and Mathematical Biology, SUPA, University of Aberdeen, Aberdeen, AB24 3UE, United Kingdom}
\address{$^2$ Institute of Medical Sciences, Foresterhill, University of Aberdeen, Aberdeen, AB25 2ZD, United Kingdom}
\eads{\mailto{cab@chrisbrackley.co.uk}, \mailto{l.ciandrini@abdn.ac.uk}, \mailto{m.romano@abdn.ac.uk}}

\begin{abstract}
The TASEP is a  paradigmatic  model from non-equilibrium statistical physics, which describes particles hopping along a lattice of discrete sites. The TASEP is applicable to a broad range of different transport systems, but does not consider the fact that in many such systems the availability of resources required for the transport is limited.  In this paper we extend the TASEP to include the effect of a limited number of \emph{two} different fundamental transport resources: the hopping particles, and the ``fuel carriers'', which provide the energy required to drive the system away from equilibrium. As as consequence, the system's dynamics are substantially affected:  a ``limited resources'' regime emerges, where the current is limited by the rate of refuelling, and the usual coexistence line between low and high particle density opens into a broad region on the phase plane. Due to the combination of a limited amount of both resources, multiple phase transitions are possible when increasing the exit rate $\beta$ for a fixed entry rate $\alpha$. This is a new feature that can only be obtained by the inclusion of both kinds of limited resources. We also show that the fluctuations in particle density in the LD and HD phases are unaffected by fluctuations in the number of loaded fuel carriers, except by the fact that when these fuel resources become limited, the particle hopping rate is severely reduced.
\end{abstract}


\maketitle

\section{Introduction}
The totally asymmetric simple exclusion process (TASEP) is one of the fundamental models of non-equilibrium statistical mechanics \cite{Schmittmann1995,Schutz2001,chou_non-equilibrium_2011}. Essentially a driven diffusion model, it has many applications in physics and beyond, including traffic models~\cite{chowdhury_physics_2005}, the movement of molecular motors in biological systems~\cite{pierobon_traffic_2009}, and protein synthesis in messenger RNA (mRNA) translation~\cite{Shaw2003, dong_towards_2007}. It also belongs to the same universality class as some surface growth models~\cite{Queiroz2008}. In this paper we study a constrained TASEP where finite resources are shared among several lattices. By \emph{finite resources} we mean a constrained number of both particles and ``fuel carriers'', whose role is to provide the energy needed to the movement of the particles. Molecular motors requiring ATP or GTP molecules are an example of such systems occurring in nature.  In this paper we introduce a new model which includes the finite availability of both resources, in contrast to previous works where the effect of having a finite number of a single type of resource was studied in isolation~\cite{Adams2008, Cook2009b, Cook2009, ciandrini_preparation_2010,Brackley2010,Brackley2010b}. As a result, multiple phase transitions can occur when varying one of the fundamental parameters of the model --the exit rate $\beta$-- while keeping the rest of the parameters constant: the system can go from a high density regime, to a shock phase, then to a high density phase again, visit the shock phase once more, and finally reach a low density phase. This is a  novel effect that emerges only by combining both limited resources. We use a mean-field approach and verify our results by means of Monte Carlo simulations.

\begin{figure}
\centering
\includegraphics{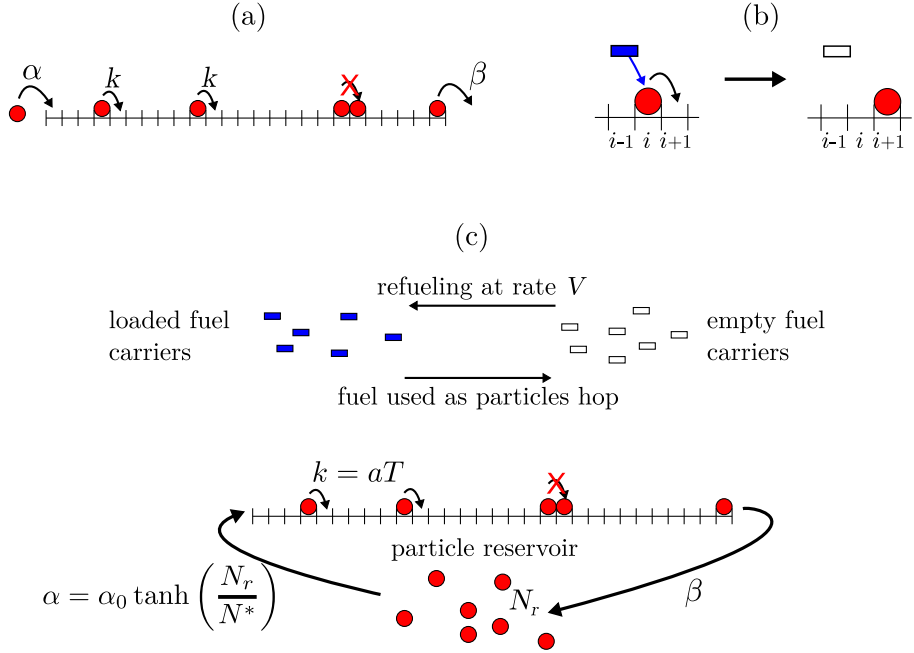}
\caption{Schematic diagrams describing the various systems. (a) The TASEP with open boundaries in its most simple form. Particles enter at fixed rate $\alpha$, hop at fixed rate $k$ and leave with rate $\beta$. (b) A finite number of fuel carriers can be introduced. When a particle hops, fuel from one carrier is used. (c) Complete model with finite fuel carrier and finite particles. The entry rate depends on the number of particles in the reservoir, and the hopping rate depends on the number of loaded fuel carriers. \label{diagram} }
\end{figure}

In its most simple form, the TASEP consists of a 1D lattice of $L$ sites upon which particles can sit, see \fref{diagram}(a). Each site can be occupied only by one particle at a time, and particles move from site to site in one direction (say rightward) with a hopping rate $k$. Since particles cannot pass each other, movement requires that the downstream site is vacant. A system with open boundaries, as we shall consider here, can display rich dynamics with multiple boundary induced phases~\cite{krug_boundary_1991}. Particles are allowed to hop onto the lattice with rate $\alpha$ at one end, and off of the lattice with rate $\beta$ at the other. For a system with constant internal hopping rate $k$ it is possible to solve the steady-state of the system exactly~\cite{Derrida1992,Schutz2001,Schutz1993,Derrida1993}, whilst the full relaxation dynamics have been solved using matrix methods~\cite{Nagy2002,deGier1999}. Mean-field methods have also been extensively used~\cite{Derrida1992}, since they are easily tractable and yield a good approximation in many cases. There has been much extension of this simple model, for example variable hopping rates (site or particle dependent)~\cite{Kolomeisky1998,Shaw2004-2,Harris2004}, extended particles which cover more than one site~\cite{Shaw2003,Lakatos2003}, branching lattices~\cite{Embley2008,Neri2011}, particles which have multiple internal states~\cite{Chowdhury2008,Klumpp2008,Ciandrini2010}, as well as a TASEP with a constrained reservoir of particles~\cite{Adams2008, Cook2009b, Cook2009, ciandrini_preparation_2010}.

For the standard TASEP we denote the occupation of the $i$th lattice site $n_i=1$ if the site contains a particle and $n_i=0$ otherwise. The system is characterised by the steady-state particle current $J$ (the rate at which particles pass any given point on the lattice), and the mean site occupancy (often called density) $\rho_i=\langle n_i \rangle$, where $\langle \cdots \rangle$ denotes average over realisations of the system (which we assume is ergodic, so this is equivalent to a time average). The average density is therefore given by $\rho=L^{-1}\sum_i \rho_i$. There are four possible phases depending on the values of $\alpha$ and $\beta$: the entry limited or low density (LD) phase, the exit limited or high density (HD) phase, a maximal current (MC) phase where the current depends only on the internal hopping rate, and a mixed LD-HD or shock phase (SP). A mean-field approach \cite{Derrida1992} (which turns out to be exact in the $L\rightarrow\infty$ limit) can be used to calculate $J$ and $\rho_i$ for given $\alpha$ and $\beta$. The density in the bulk (far from the ends of the system) is given by
\begin{equation}\label{sTden}
	\begin{array}{lll}
	\bf{(LD)}  &\rho_\mathrm{LD}=\alpha/k & \textrm{for }  \alpha<\beta, \alpha< k/2 \;,\\
	\bf{(HD)}  &\rho_\mathrm{HD}=1-\beta/k & \textrm{for }  \beta<\alpha, \beta< k/2 \;,\\
	\bf{(MC)}  &\rho_\mathrm{MC}=1/2 & \textrm{for } \beta,\alpha \geq k/2 \;,
	\end{array}
\end{equation}
and the current is always given by $J=k\rho(1-\rho)$. The SP, which occurs for $\alpha=\beta< k/2$, presents an HD region on the right of the lattice and an LD region on the left, separated by a boundary which diffuses freely through the lattice. This has often been described using a domain wall (DW) theory \cite{santen_asymmetric_2002}. Due to the free diffusion of the DW a time average of the density in SP gives $\rho_{\mathrm{SP}}=1/2$, but the current depends on the density in the LD and HD regions of the lattice, i.e. $J_{\mathrm{SP}}=\alpha(1-\alpha/k)=\beta(1-\beta/k)$.

In this paper we consider several TASEPs which share a common finite pool of both particles and fuel carriers. The entry rate, which is the same for each TASEP, depends on the availability of particles in a common pool (i.e., particles which are not involved with any lattice).  A model describing several TASEPs sharing a common pool of particles has been introduced and thoroughly studied in~\cite{Adams2008,Cook2009b,Cook2009}, where the authors use the DW theory along with known exact results. In this paper we use an alternative recent mean-field (MF) approach that allows us to simplify the calculations~\cite{ciandrini_preparation_2010}. Importantly, we combine this with a model for a finite pool of fuel carriers~\cite{Brackley2010,Brackley2010b} which, as noted above, can be viewed as carriers that provide the energy which drives the motion, i.e. allowing the particles to hop. Although we consider a fixed number of fuel carriers, we suppose that it takes a finite time to ``refuel'' them with their cargo once it has been used (figure \ref{diagram}(b) shows a schematic representation of this model).  We show that novel effects arise when both types of limited resource are  considered, e.g. multiple phase transitions can occur when varying the exit rate $\beta$.
The outline of the paper is as follows: in \sref{sec:previous} we summarise the previous results for the two models separately, before describing in \sref{sec:mf} a mean-field model for a system with both a finite pool of particles \emph{and} fuel carriers which are refuelled at a finite rate (figure \ref{diagram}(c)).
 We then interpret the mean-field model results and compare them with results from Monte Carlo simulations. Finally, in \sref{sec:flucts} we analyse the effect of both limited resources on the fluctuations in the number of particles on the lattice.

\section{Finite resources - Review of previous results}\label{sec:previous}
We first introduce and describe a system containing multiple TASEPs in which each lattice shares the same reservoir of particles; then we present the concept of fuel-carriers and the effect of a finite rate of refuelling on the exclusion process dynamics. 

\subsection{Finite number of particles}
\label{sec::particles}

In this work we analyse a system of $M$ identical lattices of length $L$. The total number of available particles is $N$, while the number of free particles in the reservoir is $N_r$. Since the lattices are identical and experience the same injection and depletion rates, we observe the same phase for each. We can write the total number of particles as
\begin{equation}
N = N_{r} + LM\rho,\label{eq::r}
\end{equation}
where $\rho$ is the density on each lattice. The entry rate of the $M$ lattices depends on the number of free particles via a saturating function
\begin{eqnarray}
\alpha &=& \alpha_0 \tanh \left[ \frac{ N_r }{N^*} \right] \nonumber \\
&= & \alpha_0 \tanh \left[ \frac{N - LM\rho}{N^*} \right],\label{eq::injection_rate}
\end{eqnarray}
where the constant $\alpha_0$ gives the entry rate in the limit $N_r\rightarrow \infty$ and is an intrinsic property of the lattices~\footnote{Equation~(\ref{eq::injection_rate}) is consistent with the function used in \cite{Adams2008, Cook2009b, Cook2009}, and is relevant, e.g. for the application to protein synthesis.} . Without loss of generality, we fix the normalisation factor $N^*$ to be $LM / 2$, i.e. the total number of particles used if all the lattices were in the MC phase. 

Throughout this paper we define the different phases according to the values of $\alpha$ and $\beta$ and the resulting density $\rho$, following~\cite{ciandrini_preparation_2010}. With this choice of nomenclature we solve equations for $\alpha$ in terms of $\alpha_0$ and $N$. Since the densities in each phase are the same as those in the standard TASEP (equations \eref{sTden}), for a given set of parameters ($\alpha_0$, $\beta$, $N$) we find the resulting $\alpha$ which determines the phase; e.g. if $\alpha<\beta$ and $\alpha<k/2$  the system will be in the LD phase. By substituting equation (\ref{eq::injection_rate}) into these inequalities, we get a representation of the different phases on the $\alpha_0\mbox{--}\beta$ plane.

\begin{figure}
\centering
\includegraphics{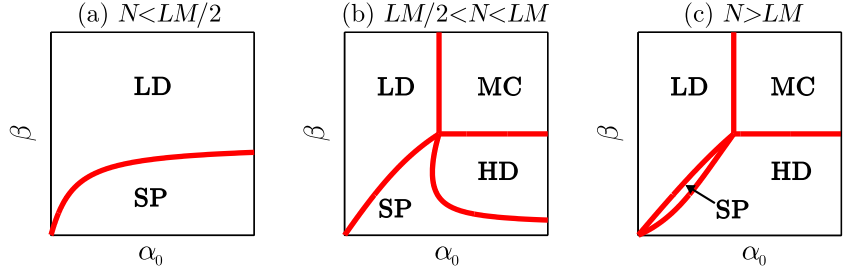}
\caption{Phase diagrams for a TASEP of length $L=500$ with a finite number of (a) $N=225$, (b) $N=475$, and (c) $N=600$ particles (infinite amount of loaded fuel carriers). For small $N$ there are not enough particles to support the HD and MC phases, and the SP (coexistence) line opens into a region. For $N>LM/2$ all four phases can be obtained. If $N$ is increased further the HD phase grows at the expense of a shrinking SP. \label{fig:FPonlyphase} }
\end{figure}

As a consequence of having a finite number of particles, we encounter different regimes for small, mid-range and large values of $N$. We show typical phase diagrams for these regimes in \fref{fig:FPonlyphase}. If $N<LM/2$, then the HD and MC phases no longer exist -- there are too few particles to support the high density or maximal current phases. Instead, there are only two phases: the LD phase and the SP (figure \ref{fig:FPonlyphase}(a)). As described in the previous section, the latter occurs when the entry and exit rates are equal, i.e.,  $\alpha=\beta$, and there is coexistence between an LD region and an HD region; since $\alpha$ depends on both $\alpha_0$ and $\beta$ (through its dependence on the bulk density), the line opens into a region on the $\alpha_0$-$\beta$ phase plane. That is to say, the condition $\alpha=\beta$ is fulfilled for a certain range of $\alpha_0$ \cite{ciandrini_preparation_2010}. If $N=LM/2$, then the lattices can support an MC phase, and for $N>LM/2$ there are enough particles for an HD phase to exist (figure \ref{fig:FPonlyphase}(b)). As $N$ is increased, the size of the HD phase on the $\alpha_0$-$\beta$ plane increases, at the cost of reducing the size of the SP phase (figure \ref{fig:FPonlyphase}(c)). For $N\gg LM$ the SP phase reduces to a line and we recover the original unconstrained TASEP. 

In the unconstrained TASEP within the SP, the LD and HD regions of the system are separated by a domain wall (DW) which can diffuse freely across the lattice. However, if there is a finite number of particles, in the case of a single lattice the DW is pinned to one position~\cite{Adams2008,Cook2009b} (actually the DW fluctuates about its mean position like a noisy damped oscillator). This is because if the DW were to move to the right this would increase the number of free particles, increasing the entry rate and therefore driving the DW leftwards. Similarly if the DW moves to the left the number of free particles decreases, decreasing the entry rate and driving the DW rightwards. The opening of the SP line into a region on the $\alpha_0$-$\beta$ plane is possible because a different mean position for the DW corresponds to a different value of $\alpha_0$, while keeping $\alpha=\beta$. Hence, the system can maintain $\alpha=\beta$ for different values of $\alpha_0$. As detailed in \cite{Cook2009}, if more than one TASEP is in contact with the same pool of particles, the DW on each lattice once again performs a random walk; there is however a pinning of the total number of particles on all lattices.

\subsection{Finite reloading time for fuel carriers}\label{sec:ffc}

In many of the systems that can be described by a driven lattice gas, the energy required for the advancement of the particles is obtained from some kind of finite resource. For instance, molecular motors consume ATP molecules, and ribosome movement on mRNAs during protein synthesis requires aminoacylated tRNA complexes and GTP. The latter case has recently been described in \cite{Brackley2010,Brackley2010b,Brackley2011}, and here we briefly review those results, before in the next section combining this with the a finite pool of particles model.

We consider a finite number $\bar{T}$ of fuel carriers, $T$ of which are carrying fuel. Every time a particle moves, the fuel from one of the loaded carriers is used, and hence, $T$ is reduced by one. The hopping rate of particles depends on the availability of loaded fuel carriers, and the empty carriers are refuelled at a rate $V$. For simplicity, the hopping rate is taken to be directly proportional to the loaded fuel carriers, i.e.
\begin{equation}
k=aT,\label{kdef}
\end{equation}
where $a$ is a constant.  Moreover, the rate of refuelling is taken to depend on the number $\bar T-T$ of unloaded fuel carriers as
\begin{equation*}
V=\frac{V_{0} (\bar{T}-T)}{b + \bar{T} -T},
\end{equation*}
which has the form of the well known Michaelis-Menten equation in biochemistry. The recharging rate is therefore a saturating function of the number of empty carriers ($\bar{T}-T$) with maximum value $V_0$ and saturation determined by the constant $b$. Any saturating function will give the same qualitative results, but the above formulation allows for a straightforward analytical treatment~\cite{Brackley2010,Brackley2010b}~\footnote{An alternative model would be to have a hopping rate which is a saturating function of $\bar{T}$, and then have a constant refuelling rate. This would give qualitatively similar behaviour to the present definitions. Our choice is most applicable to protein synthesis, i.e. refuelling due to an enzymatic reaction. }. 

\begin{figure}
\centering
\includegraphics{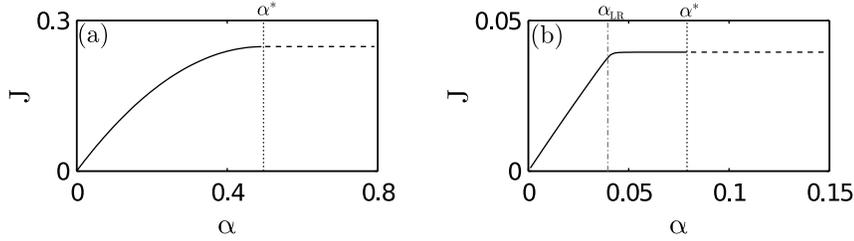}
\caption{Plots showing the current as a function of $\alpha$ for different parameters. In (a) $V_0=300~\mathrm{s}^{-1}$, and in (b) $V_0=20~\mathrm{s}^{-1}$. In both cases $a=2\times10^{-4}$, $\bar{T}=5000$, and $b=50$. Solid lines show the current in the LD phase, $J_{\mathrm{LD}}=\mathcal{J}(\alpha)$, and dashed lines the current in the MC phase, $J_{\mathrm{MC}}=\alpha^*/2$. The dotted line is at $\alpha^*$, where there is a transition from LD$\rightarrow$MC. The dot-dashed line in (b) shows the value of $\alpha_{\mathrm{LD}}$ as estimated in equation (\ref{alphaLR}). \label{fig:calJ} }
\end{figure}

In this section we describe a collection of $M$ identical TASEPs with a finite number of fuel carriers (but no constraints on the number of particles); following the common mean-field treatment~\cite{Derrida1992} the particle density on the $i$th site of each lattice is given by
\begin{equation}
\begin{array}{rclcl}
\frac{d\rho_1}{dt}&=&\alpha (1-\rho_1) - k\rho_1(1-\rho_2), &&\\
\frac{d\rho_i}{dt}&=&k\rho_{i-1} (1-\rho_i)- k\rho_i(1-\rho_{i+1}), &~& i=2,...L-1,\\
\frac{d\rho_L}{dt}&=&k\rho_{L-1} (1-\rho_L) - \beta\rho_L.&& 
\label{mnfld}
\end{array}
\end{equation}
The inclusion of a finite pool of fuel carriers leads to the additional equation
\begin{equation}
\frac{d T}{dt} = \frac{V_0 (\bar{T}-T)}{b+\bar{T}-T} - \sum_{j=1}^{(L-1) M} k\rho_j(1-\rho_j), \label{dTdt}
\end{equation}
where the sum is over all of the $L-1$ sites which use fuel carriers on each of the $M$ lattices. We assume that the particles do not require a fuel carrier to leave the $L$th site, i.e. the exit rate $\beta$ is constant. In the steady-state we identify the term under the sum in \eref{dTdt} as the particle current, and using equation \eref{kdef} we find
\begin{equation}
k=a\bar{T}-\frac{a b J (L-1)M}{V_0-J (L-1)M}, \label{eq:kofj}
\end{equation}
i.e., the hopping rate is now itself a function of the current.  Following \cite{Brackley2010,Brackley2010b}, upon solving equations \eref{mnfld} in the steady-state we find the four phases as in the original TASEP, but now the current and density are given as follows
\begin{equation}
\begin{array}{llll}
\textbf{(LD)}  & J_{\mathrm{LD}}=\mathcal{J}(\alpha), &\rho_{\mathrm{LD}}=\mathcal{D}(\alpha)
& \textrm{for} ~ \alpha<\alpha^* ~\textrm{and}~\alpha<\beta,\\
\textbf{(HD)}  & J_{\mathrm{HD}}=\mathcal{J}(\beta), &\rho_{\mathrm{HD}}=1-\mathcal{D}(\beta)
& \textrm{for} ~\beta<\alpha^* ~\textrm{and}~ \beta<\alpha,\\
\textbf{(MC)}  & J_{\mathrm{MC}}=\alpha^*/2, &\rho_{\mathrm{MC}}=1/2
& \textrm{for}  ~\alpha,\beta\geq \alpha^*, \\
\textbf{(SP)}  & J_{\mathrm{SP}}=\mathcal{J}(\alpha), &\rho_{\mathrm{SP}}=1/2
& \textrm{for} ~\alpha=\beta<\alpha^*,
\end{array}
\label{eq:MFsol}
\end{equation}
where
\begin{eqnarray*}
\fl \mathcal{J}(\alpha)=\frac{1}{2} \left[ \alpha \left( 1-\frac{\alpha}{a(\bar{T}+b)} \right) + \frac{\bar{T}}{\bar{T}+b} \frac{V_0}{L'M} \right.\nonumber \\
\left.-\sqrt{\left( \frac{\bar{T}}{\bar{T}+b}\frac{V_0}{L'M} +\alpha\left(1-\frac{\alpha}{a(\bar{T}+b)}\right) \right)^2 -\frac{4\alpha(a\bar{T}-\alpha)}{a(\bar{T}+b)}\frac{V_0}{L'M} } \right]\;, \label{eq:Jcal} \nonumber \\
\fl \mathcal{D}(\alpha)=\frac{1}{2\alpha} \left[ \alpha\left(1+\frac{\alpha}{a(\bar{T}+b)}\right) - \frac{\bar{T}}{\bar{T}+b}\frac{V_0}{L'M} \right. \nonumber \\
\left.+ \sqrt{ \left( \frac{\bar{T}}{\bar{T}+b}\frac{V_0}{L'M} +\alpha\left(1-\frac{\alpha}{a(\bar{T}+b)}\right) \right)^2 -\frac{4\alpha(a\bar{T}-\alpha)}{a(\bar{T}+b)}\frac{V_0}{L'M}  }\right]\;,\label{eq:Dcal}
\end{eqnarray*}
and
\begin{equation*}
\alpha^*=\frac{a}{4}(\bar{T}+b) +\frac{V_0}{L'M}-\sqrt{ \left(\frac{a}{4}(\bar{T}+b)+\frac{V_0}{L'M}\right)^2 -a \bar{T} \frac{V_0}{L'M}  }, \label{eq:as}
\end{equation*}
with $L'=L-1$. The behaviour of these functions as $\alpha$ and $\beta$ are varied depends on the parameters $a$, $b$ and $V_0/L'M$. By considering the steady-state of equation (\ref{dTdt}) and noticing that the maximal value that the recharging rate can possibly have is equal to $V_0$, we note that the particle current is limited from above by $V_0/L'M$. Hence, there are substantially two different cases: (i) if $V_0/L'M\gg 1$, the recharging rate is very fast and the particle current is not influenced by it; we recover the results of the original TASEP; (ii) if in contrast $V_0/L'M\ll 1$, the recharging of the fuel carriers can limit the value of the particle current. \Fref{fig:calJ} shows the current for sets of parameters corresponding to each case. In case (ii) (figure \ref{fig:calJ}(b)), $\mathcal{J}(\alpha)$ shows a sharp change from increasing with $\alpha$, to almost independent of $\alpha$ (though we note that the derivative of $\mathcal{J}(\alpha)$ remains continuous). The value of $\mathcal{J}(\alpha)$ is severely reduced compared to the one obtained for case (i) (see figure \ref{fig:calJ}(a)). We refer to the regime where the current appears independent of $\alpha$ as a limited resources (LR) regime, since the rate at which fuel is used by the particles approaches the rate at which fuel carriers are reloaded. Thus the pool of loaded carriers becomes depleted and the hopping rate $k$ reduces. For some choices of $a$ and $b$, the LR regime exists within each of the phases (LD, HD and MC). In the LR regime within the LD phase, the sensitivity of the current to changes in $\alpha$ or $\beta$ is greatly reduced, whilst the sensitivity of the density is greatly increased. In the MC phase, the current is greatly reduced in the LR regime compared to that in case (i). For further details see~\cite{Brackley2010,Brackley2010b}.

The onset of the LR regime depends particularly on the value of the three quantities $a\bar{T}$, $V_0/LM$ and $b/\bar{T}$, the former two controlling at what value of $\alpha$ or $\beta$ the onset will occur, and the latter controlling the sharpness of the change in behaviour. In the rest of this paper we choose $b$ such that there is a sharp onset of LR, and take $V_0$ as the control parameter for the fuel carriers, fixing the other parameters. This choice not only gives the most interesting dynamics, but it has also been shown to be the biological relevant regime in the context of protein synthesis~\cite{Brackley2010,Brackley2010b}. When the onset of LR is sharp we can estimate the value of $\alpha$ or $\beta$ at which this occurs by equating the rate of fuel carrier use (approximately $\alpha L'M$ for small $\alpha$ in LD and $\beta L'M$ for small $\beta$ in HD) and the maximum recharging rate. This gives
\begin{equation}
\alpha_\mathrm{LR}=\beta_\mathrm{LR} \approx\frac{\bar{T} V_0}{(\bar{T}+b)L'M}. \label{alphaLR}
\end{equation}
This value is represented in figure \ref{fig:calJ}(b) by a dot-dashed vertical line. As it is shown there, the estimation predicts quite accurately the onset of the LR regime.

\section{Constrained reservoir of particles and finite refuelling rate}\label{sec:mf}

\begin{figure}
\centering
\includegraphics{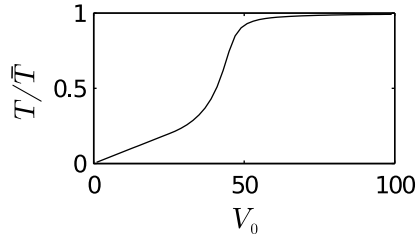}
\caption{Steady-state fuelling level $T/\bar{T}$ as a function of rate $V_0$ for a model with no constraints on the number of particles (\sref{sec:ffc}). Other parameters are $L=500$, $\alpha=0.5~\mathrm{s}^{-1}$, $\beta=0.1~\mathrm{s}^{-1}$, $a=2\times10^{-4}~\mathrm{s}^{-1}$, $b=50$, and $\bar{T}=5000$. For small $V_0$ the number of loaded fuel carriers is depleted; for large $V_0$ the carriers are practically always fully loaded, and we recover the original unconstrained TASEP. \label{fig:V0} }
\end{figure}

 A much more realistic model for natural processes such as biological transport has to include the finite availability of both particles and fuel carriers. As we show later  in this section, it is only when combining the two schemes discussed above that we can see emerging novel effects, such as multiple phase transitions. Analogous to the dependence of the entry rate $\alpha$ on the number of particles $N$, the steady-state proportion of loaded fuel carriers is a saturating function of $V_0$ (see \fref{fig:V0}). Therefore, by regarding $N$ and $V_0$ as control parameters, we can vary the number of available particles and \emph{loaded} carriers respectively; in both cases a saturating function of the resource determines the dynamics. 

The quantities of interest are the particle current $J$, the number of loaded fuel carriers $T$ and the particle density $\rho$, which is linked to the number of free particles by equation \eref{eq::r}. The expressions for $J$ and $\rho$ given in equations (\ref{eq:MFsol}) still hold; however $\alpha$ is no longer a control parameter, and it can be eliminated using equation \eref{eq::injection_rate}. This is the case in all phases except in the SP: since the DW cannot move freely on the lattice, the mean density now depends on the size of the LD and HD regions. We calculate the average value $\rho_\mathrm{SP}$ using equation (\ref{eq::injection_rate}); using the fact that 
in this phase $\alpha=\beta$ leads to
\begin{equation}
	\rho_{\mathrm{SP}}=\frac{N}{LM} \left( 1-\frac{N^*}{N}  \tanh^{-1}\left( \frac{\beta}{\alpha_0} \right)\right).
	\label{eq:rhoSP}
\end{equation}
The hopping rate $k$ (and therefore $T$) in each phase can be found by substituting the appropriate equation for the current $J$ in \eref{eq:kofj}.

\begin{figure*}
\centering
\includegraphics{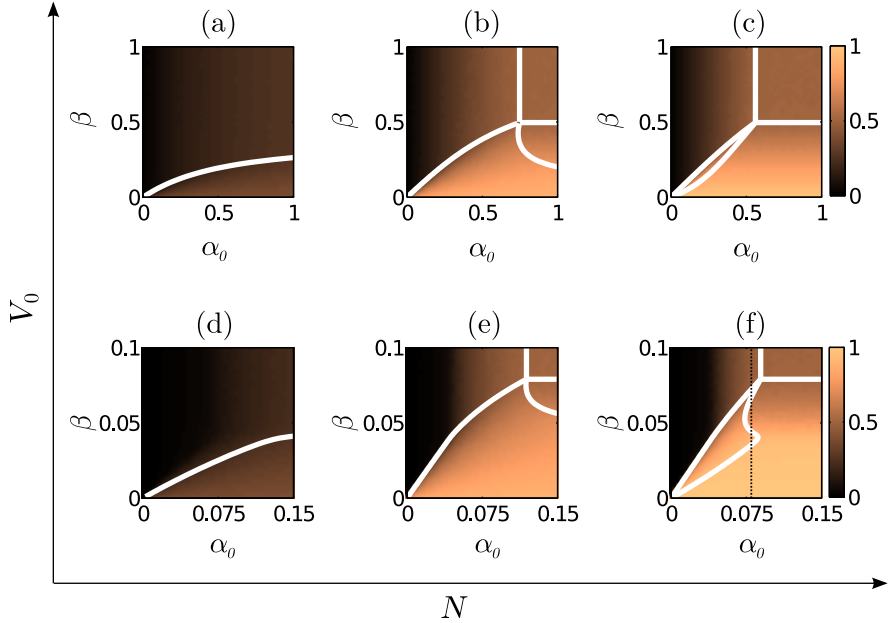}
\caption{Colour on-line. Phase diagrams in the $\alpha_0$-$\beta$ plane at different values of $V_0$ and $N$. White lines show the phase boundaries as determined by the mean-field model of \sref{sec:mf}. Colour maps show the steady-state particle density $\rho$ from Monte Carlo simulations. Parameters used are $V_0=20$ or $300~\mathrm{s}^{-1}$, and $N=200,~450$, or $600$. Other parameters are $L=500$, $M=1$, $a=2\times10^{-4}~\mathrm{s}^{-1}$, $b=50$, and $\bar{T}=5000$. The dotted line in (f) shows the line of constant $\alpha_0$ used in figure \ref{fig:HDSPHDLD}. \label{fig:bigphase} }
\end{figure*}

We now turn to the problem of finding the boundaries between the different phases as functions of $\alpha_0$, $\beta$, $N$ and $V_0$ --- i.e., eliminating $\alpha$. Our aim is to draw the $\alpha_0\mbox{-}\beta$ phase plane for any given values of $V_0$ and $N$. As noted in previous sections, due to finite particles, the SP line opens into a region, and the HD and MC phases do not exist if $N<LM/2$. Each phase boundary can be written in terms of either $\alpha_0$ as a function of $\beta$, or vice versa.  We now consider the boundaries between each phase in turn, consulting \fref{fig:FPonlyphase} as an ansatz for the arrangement of the phases.
\begin{description}
\item \emph{(i) MC/LD phase boundary.} The MC phase can exist if $N>LM/2$. If we consider starting in the MC phase with large $\beta$ and reducing $\alpha_0$, we will cross into the LD phase when $\alpha=\alpha^*$. Using \eref{eq::injection_rate} this gives an equation 
for the boundary in the $\alpha_0$-$\beta$ plane, where
\begin{equation*}
\alpha_0= \alpha^* \coth \left( \frac{N}{N^*} - \frac{LM}{2 N^*} \right) ~ \mathrm{for} ~~\beta\geq\alpha^* ~~\mathrm{and}~~N>LM/2\;.
\end{equation*}
Note that $\alpha^*$ depends on $a$, $b$, $\bar T$, $V_0$, $L$ and $M$. Hence the MC/LD boundary is a vertical line on the $\alpha_0\mbox{-}\beta$ phase plane.

\item \emph{(ii) MC/HD phase boundary.} If we now consider starting in the MC phase with large $\alpha_0$ and reducing $\beta$, then we cross into the HD phase when $\beta=\alpha^*$, i.e., in the $\alpha_0$-$\beta$ plane this boundary is given by the horizontal line 
\begin{equation*}
\beta= \alpha^* ~ \mathrm{for} ~~ \alpha_0\geq\alpha^* \coth \left( \frac{N}{N^*} - \frac{LM}{2 N^*} \right) ~~\mathrm{and}~~ N>LM/2\;.
\end{equation*}

\item \emph{(iii) HD/SP phase boundary.} Here we consider moving from the HD to the SP. For the system to be in the HD phase we require $\beta<\alpha^*$, $N>LM/2$ and $\beta<\alpha$. Using equation \eref{eq::injection_rate} in the latter inequality gives $\beta< \alpha_0 \tanh (N_r/N^*)$; using equation \eref{eq::r} and the density in HD phase gives the following equation for the boundary 
\begin{equation}
\alpha_0=\beta \coth \left(  \frac{N}{N^*}  - \frac{LM}{N^*}(1-\mathcal{D}(\beta))  \right) ~ \mathrm{for} ~~ \beta<\alpha^* ~~\mathrm{and}~~ N>LM/2\;. \label{eq:HDSP}
\end{equation}
Hence, the HD/SP phase boundary is a curved line on the $\alpha_0\mbox{-}\beta$ phase plane.

\item \emph{(iv) LD/SP phase boundary.} Finally we consider starting in the LD phase and moving to the SP. In LD we require $\alpha<\beta$ and $\alpha<\alpha^*$. Finding an expression for these inequalities poses some difficulty, since by using equation \eref{eq::injection_rate} and the density in the LD phase we obtain
\begin{equation*}
\alpha=\alpha_0 \tanh \left( \frac{N}{N^*} - \frac{LM}{N^*}\mathcal{D}(\alpha) \right),
\end{equation*}
an equation which cannot be solved analytically to find $\alpha$ as a function of $\alpha_0$. Instead we solve this numerically, setting $\alpha=\beta$ (which occurs at the SP) to find $\beta$ as a function of $\alpha_0$. This gives another curved line on the $\alpha_0\mbox{-}\beta$ phase plane.
\end{description}
We can then construct the phase plane by plotting 
the phase boundaries ($\beta$ as a function of $\alpha_0$) for any given values of $N$ and $V_0$. Unless otherwise stated, throughout the rest of this paper we use parameters $\bar{T}=5000$ so that $\bar{T} \gg LM$ and hence, it is always the refuelling which is the limiting process, and not the total number of fuel carriers. This represents realistic scenarios in biological transport processes, such as protein synthesis.
We set the time scale of the system by choosing $a=2\times10^{-4}~\mathrm{s}^{-1}$, such that the maximum hopping rate is $k=1~\mathrm{s}^{-1}$. A value of $b=50$ then gives a sharp onset of LR as shown in \fref{fig:calJ}(b). The phase diagram boundaries calculated using the mean-field approach are shown in \fref{fig:bigphase} using white lines.

To test the validity of the mean-field results derived above we perform simulations using a continuous time Monte Carlo method \cite{Bortz1975}. The length of the Monte Carlo time step is chosen from an exponential distribution, such that the events occur according to a Poisson process, with a single event occurring at each step. Possible events are the movement of a particle (either on to, along, or off of the lattice) or the refuelling of a fuel carrier. The event which occurs is chosen stochastically from the set of particles which have a vacancy to their right and the set of empty fuel carriers. Particles are chosen with a probability such that they move with a rate $k$, and empty fuel carriers are chosen with a probability such that they are recharged with rate $V$; after each event $T$ is updated accordingly.  To remove any transient effects associated with the initial condition we disregard the first $5\times10^6$ time steps. Assuming that the system is ergodic we average currents and densities over at least a further $4\times10^7$ steps.

In \fref{fig:bigphase} we present a series of phase planes for different values of $N$ and $V_0$. We show the particle density $\rho$ obtained from Monte Carlo simulations as a colour map with the phase boundaries from the mean-field model overlaid (white lines). Note that the mean-field model very closely predicts the boundaries. Here we show data for $M=1$ lattices, but the plots look the same for $M>1$ with appropriately scaled parameters. Introducing more lattices does not change the macroscopic behaviour, but changes the microscopic behaviour for the SP (see \sref{sec:mf_shock}). 

\begin{figure}
\centering
\includegraphics{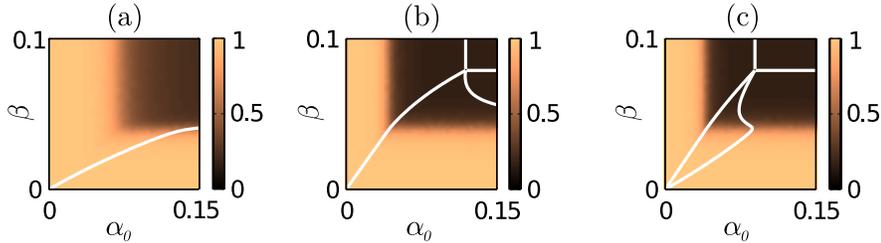}
\caption{Colour on-line.  Colour maps showing Monte Carlo results for the steady-state fuelling level $T/\bar{T}$ at different values of $\alpha_0$ and $\beta$ for small $V_0=20~\mathrm{s}^{-1}$. White lines show phase boundaries as predicted by the mean-field model. (a) Small $N=200$; (b) mid-range $N=450$; (c) large $N=600$. For these parameters we have $\alpha_\mathrm{LR}=\beta_\mathrm{LR}\approx0.040$ from \eref{alphaLR}. Other parameters are as \fref{fig:bigphase}. \label{fig:phase2} }
\end{figure}

In \fref{fig:bigphase} we note that for small $V_0$ the phase diagrams look broadly similar to the large $V_0$ case, but as we expect from ~\cite{Brackley2010,Brackley2010b}, the phase transitions occur at much smaller values of $\alpha_0$ and $\beta$. We also obtain a limited resources (LR) regime within each of the phases. This can seen in figure \ref{fig:phase2}, which shows colour plots for the proportion of charged fuel carriers ($T/\bar{T}$) for different values of $N$, with small $V_0$. The onset of the LR regime can be clearly seen as $T/\bar{T}$ decreases dramatically over a small range of $\alpha_0$ or $\beta$. For large $V_0$ the refuelling is so quick that $T/\bar{T}$ is constant through all phases, i.e. we recover the results for a TASEP with a finite pool of particles, but no constraints on the fuel carriers, since these are refuelled almost as soon as they are used (data not shown). For clarity, throughout the rest of the paper when we refer to LR or limited resources, we specifically mean the regime where the pool of loaded fuel carriers has become depleted.

Crucially, the presence of the LR regime also alters the shape of the phase boundaries; a noticeable ``kink'' can be seen in the LD/SP and HD/SP phase boundaries at the point of LR onset. Particularly striking is the shape of the HD/SP boundary for the large $N$ small $V_0$ case, and we examine this in detail in \sref{sec:multiple}.

\subsection{The Shock Phase}\label{sec:mf_shock}

As noted above, due to the finite number of particles, the coexistence line present in the original TASEP -- the shock phase -- opens out into a region on the phase plane. We examine the behaviour in this phase by locating the domain wall (DW) which separates the regions of LD and HD, and examining how the position of this is affected by the finite number of fuel carriers and how it changes at different values of $\alpha_0$ and $\beta$. The introduction of a finite number of particles also gives a change from a DW which can wander freely along the lattice, to one where the wandering is constrained by the presence of the reservoir. This is most easily explained in the case of $M=1$ lattices, where the DW is on average fixed in position. A fluctuation which leads to movement of the DW away from its mean position will change the number of particles in the reservoir; this in turn changes the entry rate $\alpha$, acting like a restoring force on the DW. In actual fact the DW executes Gaussian fluctuations about its mean value, and we discuss fluctuations further in \sref{sec:flucts}.  If multiple lattice are introduced ($M>1$), then as in \cite{Cook2009} the DW is not pinned, but rather it is the sum of the DW position on all lattices which executes Gaussian fluctuations about a mean. We focus on the $M=1$ case for the rest of this section.

The relative mean position $x \in [0,1]$ of the pinned DW (where $xL$ gives the site at which it is located) can be estimated from the mean-field model by approximating the density in the SP as follows
\begin{equation*}
\rho_{\mathrm{SP}}=x\rho_{\mathrm{LD}} + (1-x) \rho_{\mathrm{HD}}.
\end{equation*}
Since $x$ is always selected such as to maintain the condition $\alpha=\beta$, we can use equation \eref{eq:rhoSP} and the densities from \eref{eq:MFsol} to show that
\begin{equation*}
x=\frac{1}{1-2\mathcal{D}(\beta)} \left[ 1-\mathcal{D}(\beta) -\frac{N}{LM} \left(1-\frac{N^*}{N} \tanh^{-1} \left( \frac{\beta}{\alpha_0} \right) \right)\right]. \label{eq:x}
\end{equation*}
We also note that the difference between $\rho_{\mathrm{LD}} $ and $\rho_{\mathrm{HD}}$ decreases as $\beta$ increases, i.e., the ``height'' of the wall decreases.

\begin{figure}
\centering
\includegraphics{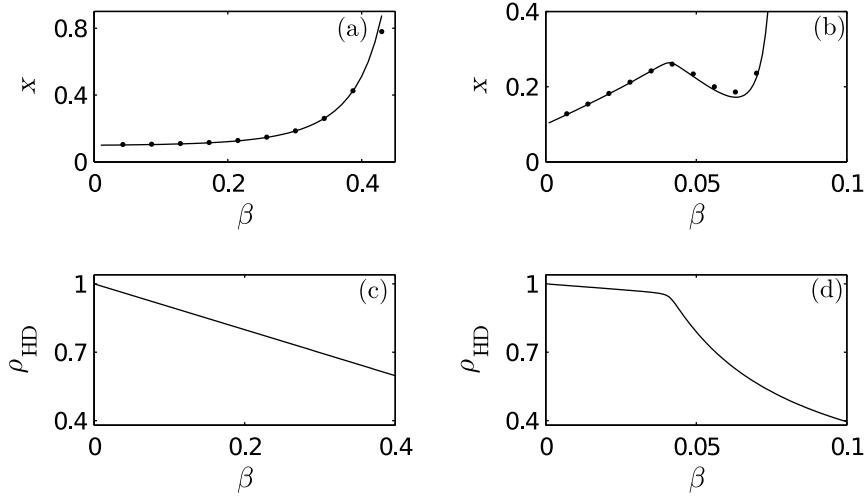}
\caption{Plots (a) and (b) show the relative mean DW position for systems in the SP with $V_0=300~\mathrm{s}^{-1}$ and $V_0=20~\mathrm{s}^{-1}$ respectively. In both cases $N=450$; in (a) $\alpha=0.6~\mathrm{s}^{-1}$, and in (b) $\alpha=0.1125~\mathrm{s}^{-1}$. In (b) the onset of the LR regime is at $\beta\approx0.04~\mathrm{s}^{-1}$. Plots (c) and (d) show how the density in the HD phase $\rho_{\mathrm{HD}}=1-\mathcal{D}(\beta)$ varies with $\beta$, again for $V_0=300~\mathrm{s}^{-1}$ and $V_0=20~\mathrm{s}^{-1}$ respectively. In the SP this is the density to the right of the DW. In (d), initially the density decreases slowly with increasing $\beta$; at the onset of LR the rate of density decrease becomes more severe - a change in the behaviour not seen in a model with an infinite number of fuel carries. \label{fig:HDden} }
\end{figure}

As it is the most interesting case, we focus on parameters where the SP has the largest area on the phase diagram, namely the mid-range $N$ cases, i.e, figures \ref{fig:bigphase}(b) and (e).  In \fref{fig:HDden} we show plots for the mean position of the DW as a function of $\beta$, for large and small values of $V_0$. Also shown for each case is the density in the HD region of the lattice (to the right of the DW) as a function of $\beta$, which will aid in the following discussion.

For large $V_0$ (figures \ref{fig:HDden} (a) and (c)) there is a monotonic increase in $x$ with $\beta$. A larger value of $\beta$ requires that more particles are present in the reservoir in order to achieve $\alpha=\beta$. At small $\beta$, $x$ is approximately constant with $\beta$; this is because the decrease in density on the HD side of the lattice causes a sufficient release of particles to maintain $\alpha=\beta$. Due to the saturating form of the function $\alpha(N_r)$ (equation (\ref{eq::injection_rate})), for larger values of $N_r$ a greater increase in $N_r$ is required to give the same increase in $\alpha$. Hence for larger values of $\beta$, the change in the HD density as $\beta$ increases no-longer releases sufficient particles to keep $\alpha=\beta$; the DW also must move towards the right such that there is a steep increase of $x$ with $\beta$.

In the small $V_0$ case there is an LR regime within the SP which results in an interesting dependence of $x$ on $\beta$ (figure.~\ref{fig:HDden}(b)); in contrast to the large $V_0$ value case, $x$ does not increase monotonically with $\beta$. We can understand this behaviour by again considering the density in the HD region of the lattice. We note that for the small $V_0$ case, $\rho_{\mathrm{HD}}$ changes differently with $\beta$ depending on whether the system is in the LR regime or not, and that the maximum in $x$ at $\beta\approx0.04~\mathrm{s}^{-1}$ corresponds to the onset of the LR regime. For $\beta<0.04~\mathrm{s}^{-1}$ we see from \fref{fig:HDden}(d) that, increasing $\beta$ results in a decrease in the density in the HD region -- and therefore a release of particles to the reservoir and an increase in $\alpha$. However the decrease of $\rho_{\mathrm{HD}}$ with $\beta$ is not enough to maintain $\alpha=\beta$. The DW must also move rightwards, i.e. there is an initial increase of $x$ with $\beta$. After the onset of LR, $\beta>0.04~\mathrm{s}^{-1}$, \fref{fig:HDden}(d) shows that $\rho_{\mathrm{HD}}$ decreases much more quickly with increasing $\beta$. So now the density on the HD side of the DW decreases much more rapidly as $\beta$ increases. The resulting release of particles would be too great to maintain $\alpha=\beta$ if the wall did not also move leftwards -- $x $ decreases again. 

Deeper within the LR regime \fref{fig:HDden}(d) we have the same situation as before: due to the saturating function $\alpha(N_r)$, for large $N_r$ we need a greater increase in $N_r$ to give the same increase in $\alpha$. The wall has to move rightward as $\beta$ increases in order to release enough particles to maintain $\alpha=\beta$. 

\subsection{Multiple Phase Transitions}\label{sec:multiple}

\begin{figure*}
\centering
\includegraphics{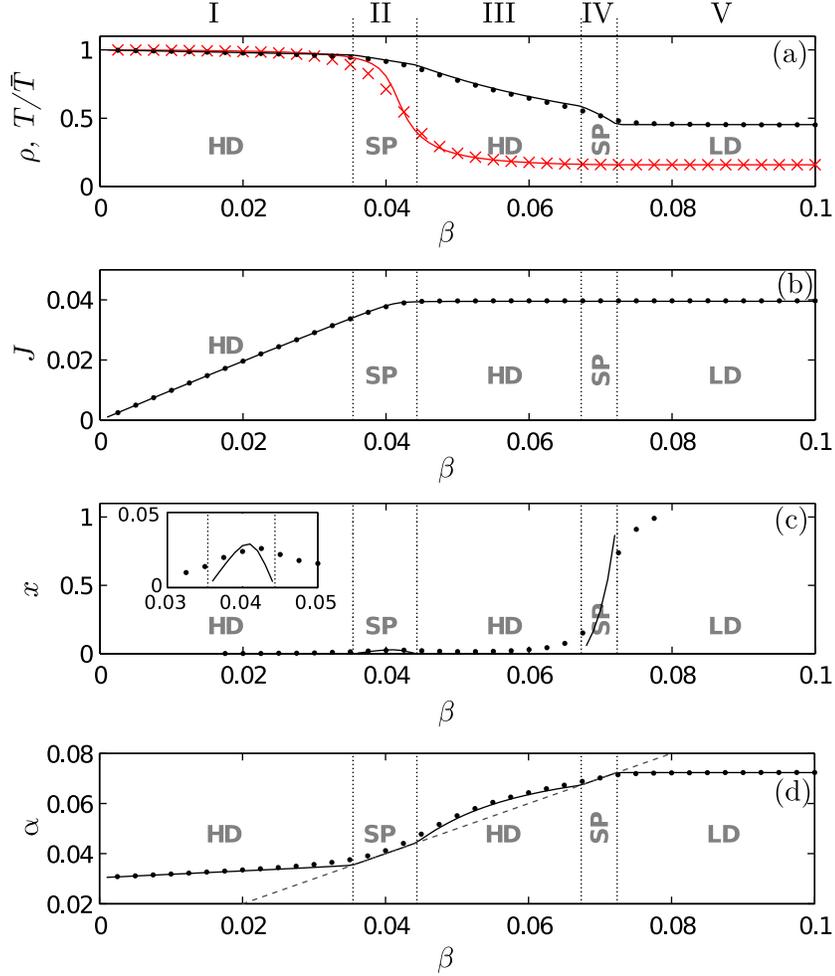}
\caption{Colour on-line. Plot showing (a) how the density $\rho$ (points) and fuelling level $T/\bar{T}$ (crosses), (b) the current $J$, (c) the mean DW position $x$, and (d) the entry rate $\alpha$ as given by equation (\ref{eq::injection_rate}), vary with $\beta$ for small $V_0$ and large $N$ at a fixed value of $\alpha_0=0.08~\mathrm{s}^{-1}$. Points show Monte Carlo results and lines the prediction from the mean-field model. The system passes through several phases and we indicate with dotted lines the positions of the boundaries as predicted by the mean-field model. We label each phase with roman numerals I-V. The inset in plot (c) shows a zoom around the SP phase II. The dashed line in (d) shows $\alpha=\beta$. \label{fig:HDSPHDLD} }
\end{figure*}

By combining the effects of both types of limited resources, we obtain a novel phase diagram (\fref{fig:bigphase}(f) and \fref{fig:phase2}(c)). There is an unusual kink shape in the phase boundary between the HD and SP regimes, given by equation \eref{eq:HDSP}. As can be seen in \fref{fig:phase2}(c), this is at the point where the system enters the limited resources regime. It is possible to draw a vertical line at constant $\alpha_0$ through the phase diagram (dashed line in figure \ref{fig:bigphase}(f)), which cuts through the phases HD$\rightarrow$SP$\rightarrow$HD$\rightarrow$SP$\rightarrow$LD as $\beta$ increases, i.e., by varying only one parameter we can go from the HD phase through a transition to SP, and then a transition back to HD, etc. \Fref{fig:HDSPHDLD} shows how the quantities $\rho$, $T/\bar{T}$, $J$, $x$ (where applicable) and $\alpha$, vary along this line of constant $\alpha_0$; we show both Monte Carlo results and the prediction of the mean-field model. The mean-field model performs well deep within each phase, but begins to show some discrepancy near the phase boundaries; we discuss this further below.

We label the phases shown in \fref{fig:HDSPHDLD} with roman numerals I--V and explain each in turn. These are that phases which are crossed by the dashed line in figure \ref{fig:bigphase}(f).
\begin{description}
\item[Phase I] At very small values of $\beta$ we have $\alpha\gg\beta$, so the system is in the HD phase. Again considering how $\rho_{\mathrm{HD}}$ varies with $\beta$, from \fref{fig:HDden}(d) we see that for small $\beta$ the slope is small, $d\rho_\mathrm{HD}/d\beta\sim-1$. A decreasing density means an increasing number of free particles, i.e., $dN_r/d\beta\sim1$; the hyperbolic tangent form of equation \eref{eq::injection_rate} means that $\alpha$ increases with $\beta$ but at a very low rate ($d\alpha/d\beta\ll1$, see figure \fref{fig:HDSPHDLD}(d)) \footnote{In the model with finite resources $\alpha$ depends on $\beta$, in contrast to the standard TASEP.}.

\item[Phase II] We arrive at phase II as follows: in phase I we started with very small values of $\beta$ such that $\alpha \gg \beta$.  By increasing $\beta$, particles are freed and therefore, $\alpha$ also increases. However, $d\alpha/d\beta \ll 1$ in phase I, and hence, we eventually reach $\alpha=\beta$, and there is a transition to an SP -- phase II. Here we have coexistence of both LD and HD separated by a DW.

The current --- and therefore the fuel carrier use rate --- increases with $\beta$ through phase I, and initially in phase II (\fref{fig:HDSPHDLD}(b)). About half way through phase II ($\beta\approx0.04~\mathrm{s}^{-1}$) the system enters the LR regime (see crosses in \fref{fig:HDSPHDLD}(a)). 

From \fref{fig:HDden}(d) we know that in the first half of phase II, the density in the HD part of the system decreases slowly with $\beta$ --- $d\rho_{\mathrm{HD}}/d\beta\sim-1$. The corresponding increase in the number of free particles would not be enough to keep $\alpha=\beta$, so the DW also moves rightward, i.e. there is an initial increase in $x$ in phase II.

At the onset of LR the slope of the curve in \fref{fig:HDden}(d) gets steeper, i.e., $d\rho_{\mathrm{HD}}/d\beta\ll-1$. As $\beta$ is increased further (in the second half of phase II) the DW must move leftwards again in order to keep $\alpha=\beta$; i.e. after initially increasing with $\beta$, $x$ then decreases as LR onsets, as shown in \fref{fig:HDSPHDLD}(c) (see inset).

\item[Phase III] Once the DW reaches the leftmost side of the lattice the system can no longer maintain the condition $\alpha=\beta$, so there is a phase transition and we re-enter the HD phase. Further increase of $\beta$ increases the number of free particles $N_r$, however since $\alpha$ is a saturating function of $N_r$, $d\alpha/d\beta$ begins to decrease. That is, as $\beta$ increases through phase III the slope of $\alpha(\beta)$ gets shallower (\fref{fig:HDSPHDLD}(d)).

\item[Phase IV] If we keep increasing $\beta$, we reach $\beta=\alpha$, and then the system enters the SP for a second time. From \fref{fig:HDden}(d) we see that deep within the LR regime $\rho_{\mathrm{HD}}$ again varies slowly with $\beta$, and the difference between the LD and HD densities is small; therefore in this second SP, changes in the density in the two regions of the lattice would not significantly change the number of free particles. Thus rapid variation of the DW position is required as $\beta$ increases in order to maintain $\alpha=\beta$.

\item[Phase V] Once the DW reaches the rightmost edge of the lattice, the system can no longer maintain the condition $\alpha=\beta$ by moving the wall, and there is a transition to the LD phase.
\end{description}

The above description accounts for the changes in DW position, current and density predicted by the mean-field theory, but as we noted previously there is some discrepancy with the Monte Carlo results, particularly near the transitions. This is due to the fact that our mean-field model assumes that the density is constant throughout the lattice, when in fact there is some change near the edges \cite{dong_towards_2007}. Also the mean-field treatment ignores correlations in the density which occur near the DW. These edge effects become less significant as $L$ is increased, and therefore the discrepancy between the mean-field and simulation results reduces (data not shown).

We also note that there is some difficulty in determining the existence and position of the domain wall. In \fref{fig:HDSPHDLD}(c) we define DW position from simulations by considering the mean particle density at each lattice site $\rho_i$; we define the existence of a DW if for any pair of adjacent lattice sites $i,i+1$ the density cuts through 0.5. Then, the position of the DW is given by lattice site $i$. The difficulty arises in the fact that this can also occur near the edges of the system when it is not in the SP. This explains why it appears that there are DWs when the system is not in the SP -- we are actually detecting the decrease in the density at the edge of the system.

\section{Fluctuations}\label{sec:flucts}

In this section we examine fluctuations in the density of particles on the lattice, focusing on the case of $M=1$ lattice. We obtain power spectra for the fluctuations in density by taking the average of the Fourier transform of several different time series \cite{Pierobon}. These time series are generated by recording the densities at time intervals of $\sim 125~\mathrm{s}$. Since our simulation method does not advance time in regular steps, the time intervals are not exactly uniform; therefore we use cubic spline fitting to obtain a time series at regular intervals. Defining $\rho(t)$ as the instantaneous density at time $t$, the power spectrum is given by $I(\omega)=\langle | \textrm{FT}\left[\rho(t)-\rho\right] |^2 \rangle$, where the angled brackets denote average over different simulation runs, and $\mathrm{FT}[f(t)](\omega)$ is the discrete Fourier transform of the time series $f(t)$. As before $\rho$ denotes the time average density. From now on, we use the following notation: if we mean the instantaneous value of a quantity at time $t$, then explicit time dependence is indicated; symbols without time dependence denote the time average of the quantity.

\subsection{LD and HD Phase}

\begin{figure}
\centering
\includegraphics{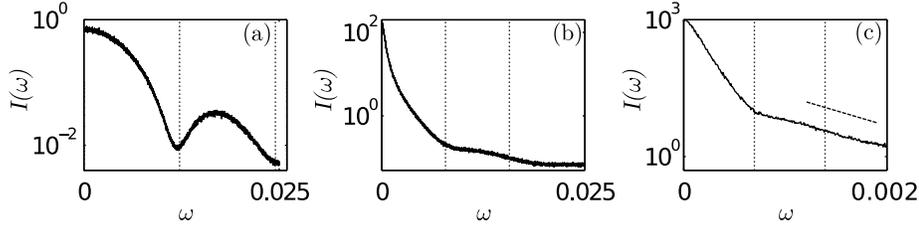}
\caption{Plots showing power spectra of the total density from simulations of a system with finite fuel carriers with $V_0=20~\mathrm{s}^{-1}$, but no constraints on the number of particles ($N\rightarrow\infty$). Plot (a) is for $\alpha=0.01~\mathrm{s}^{-1}$, (b) for $\alpha=0.04~\mathrm{s}^{-1}$, and (c) $\alpha=0.06~\mathrm{s}^{-1}$; in the latter two plots the system is in the LR regime, and we note that the oscillations have been damped out (see text). The dashed line has a slope $\omega^{-2}$, and the dotted lines in each plot are at $\omega=2\pi v/L$ and $4\pi v/L$. These spectra were obtained by averaging the Fourier transforms of 500 different time series. \label{fig:fluct} }
\end{figure}

A good approximation to the power spectrum of the density fluctuations in the LD phase for the original TASEP (no particle or fuel carriers constraints) can be found via a continuum description, with fluctuations travelling through the lattice with a velocity $v$ and an effective diffusion coefficient $D$~\cite{Adams2007}.\footnote{Due to particle-hole symmetry (which is maintained in the present model), our understanding of fluctuations in the LD phase can also be applied to the HD phase.} At low frequencies ($\omega\ll v^2/D$) the power spectrum shows oscillations, or dips in power, at unit multiple of $2\pi v/L$, i.e., at frequencies corresponding to the length of time it takes a fluctuation to traverse the entire lattice. These oscillations are damped for $\omega>v^3/DL$, after which $I\propto \omega^{-2}$; for large frequencies $I\propto \omega^{-3/2}$. For the case of finite re-fuelling, but no particle constraints (i.e. the $N\rightarrow\infty$ limit), we obtain a similar power spectrum . As shown in \fref{fig:fluct}(a), we have oscillations with dips at multiples of $2\pi v/L$ (with $v\equiv k(1-2\rho)$, see \cite{Adams2007}); at larger values of $\alpha$, when the system is in the LR regime, the oscillations are severely damped, \fref{fig:fluct}(b). This is because the hopping rate $k$ is highly reduced in the LR regime, meaning that $v$ is reduced, and hence, the damping occurs at much lower frequencies. What remains of the dips is still clearly visible in the power spectrum at low multiples of $2\pi v/L$, implying that the density fluctuations are not largely affected by the fluctuations in the hopping rate in this regime. The $I\propto \omega^{-2}$ for mid-ranged $\omega$ and $I\propto \omega^{-2/3}$ for large $\omega$ relationships are maintained even for parameters such that the mean number of loaded fuel carriers $T$ is small, conditions under which one might have though that fluctuations in $T$ would become important (data not shown). We find that actually fluctuations in loaded carrier levels become small when the mean value is small.\footnote{In fact we find that $T\approx\langle (T(t)-T)^2 \rangle$ for small $T$ in the LR. The probability of finding $T$ loaded fuel carriers appears to be very close to a Poisson distribution in this regime, but further discussion of this is beyond the scope of the present work.}

In summary, just as in the original TASEP, when there are constrained fuel carriers the oscillations in the power spectrum of the number of particles on the lattice are damped for $\omega>v^3/DL$. In the original TASEP this is most noticeable at larger values of $\alpha$, which give large values of $\rho$, and therefore small $v$. With constrained fuel carriers small $v$ is obtained in the LR regime due to the reduction in the value of $k$, i.e., at much smaller values of $\alpha$.

\begin{figure}
\centering
\includegraphics{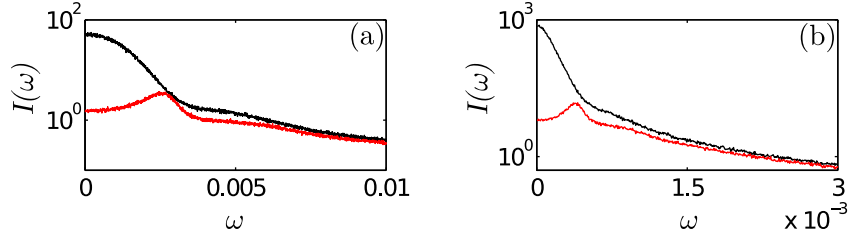}
\caption{Colour on-line. Plots comparing power spectra of density fluctuations for systems with a large number of particles (black) and a small number of particles (red). (a) Systems not in the LR regime, with $V_0=300~\mathrm{s}^{-1}$. Black lines show results for a large number of particles: $N=600$ and $\alpha_0=0.4~\mathrm{s}^{-1}$. Red lines show results for a small number of particles: $N=220$ and $\alpha_0=2.83~\mathrm{s}^{-1}$ . (b) Systems in the LR regime, with $V_0=20~\mathrm{s}^{-1}$. Again black lines show results for a large number of particles: $N=600$ and $\alpha_0=0.07~\mathrm{s}^{-1}$ . Red lines show results for a small number of particles: $N=220$ and $\alpha_0=0.65~\mathrm{s}^{-1}$ . The different values of $\alpha_0$ used in each case are chosen such that the systems we compare have the same mean density and mean hopping rate. \label{fig:fluct2} }
\end{figure}

Density fluctuations have also previously been studied for the TASEP with constrained particles~\cite{Cook2010}. The effect in the LD regime is to suppress the fluctuations. This is due to the feedback effect of the particle pool which stabilises the density (an increase in density leads to a decrease in entry rate). The effect is reduced for higher frequencies, as these correspond to short time scales over which the feedback from the particle pool has less influence, i.e. for time scales shorter than the time it takes a fluctuation to traverse the entire lattice. Turning to the present model with constrained particles \emph{and} fuel carriers, \fref{fig:fluct2} shows power spectra for systems with different total numbers of particles, both when there are no limited fuel resources (\fref{fig:fluct2}(a)) and when there are (\fref{fig:fluct2}(b)). Note that to fairly compare fluctuations from two simulations the mean densities and hopping rates (and therefore positions of the ``dips'' in the power spectrum) must be the same; we therefore choose different values of $\alpha_0$ to give the same value of $v$ in each case. Again we see little qualitative difference between the non-LR and LR cases, i.e., fluctuations in the number of loaded fuel carriers have little effect on the fluctuations in the density.

\subsection{Shock Phase}

We now consider fluctuations in the density, and in the position of the DW, when the system is in the SP. We again consider a single lattice ($M=1$), and focus on the case of small $V_0$, and a mid-range value of $N$, corresponding to figures \ref{fig:bigphase}(e) and \ref{fig:phase2}(b), for which the SP has the largest area on the $\alpha_0\mbox{-}\beta$ plane.

To locate the DW we average the occupation of each site over a short time window of length $\tau$, so as to average over short time microscopic fluctuations in the occupation~\cite{Embley2008}. The value of $\tau$ is chosen short enough so as to probe the movement of the DW on mesoscopic time scales rather than probing only the mean-field density profile. Following the approach of \cite{Embley2008} we choose for $\tau$ the smallest value which gives exactly one micro-domain wall ($\mu$DW) in the averaged profile. A $\mu$DW is defined as any point where the density crosses 0.5 (from above or below) between one lattice site and the next.

\begin{figure*}
\centering
\includegraphics{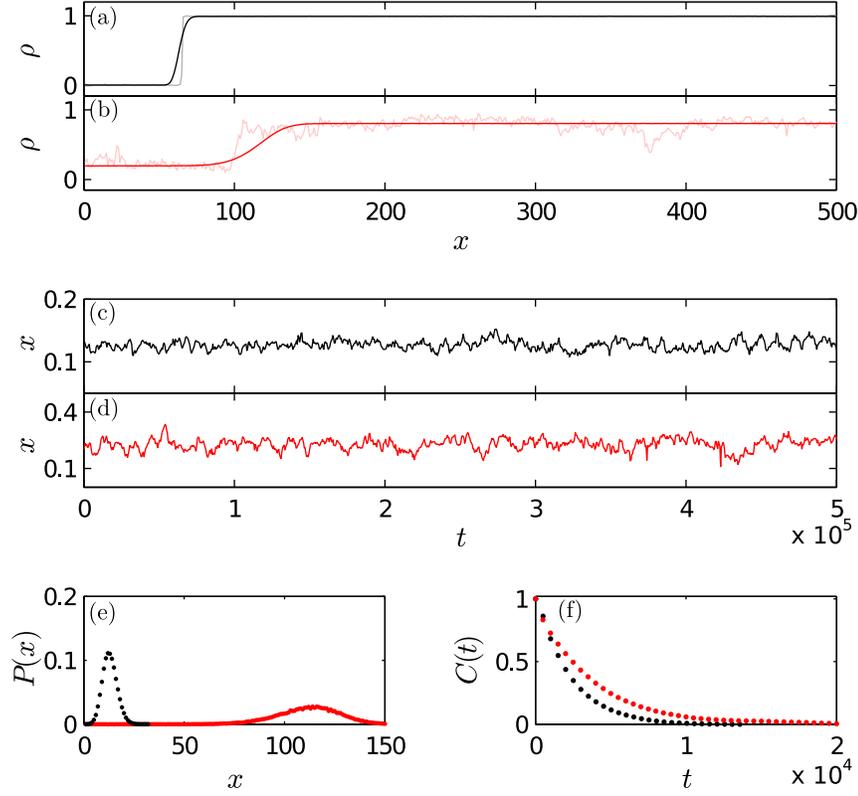}
\caption{Colour on-line. Simulation data showing the behaviour of the domain wall in the SP for small $V_0$ and mid-range $N$ corresponding to \fref{fig:bigphase}(e), for $\alpha_0=0.1125~\mathrm{s}^{-1}$ and various values of $\beta$. Plots (a)-(b) show density profiles, and (c)-(d) the DW position time courses for $\beta=0.007~\mathrm{s}^{-1}$ (black) and $\beta=0.049~\mathrm{s}^{-1}$ (red). The latter corresponds to a system which is in the LR regime. For the density profiles long time averages are shown with heavy lines, and mesoscopic time averages (from which we determine the instantaneous wall positions) with light lines. The mesoscopic averaging time was $\tau=500~\mathrm{s}$. Plot (e) shows DW position histograms (normalised) for the same two values of $\beta$, and plot (f) shows the time correlation function from \eref{tcor}. 
 \label{fig:SPmeanX} }
\end{figure*}

We show in \fref{fig:SPmeanX} the behaviour of the DW at different values of $\alpha_0$ and $\beta$. Figures \ref{fig:SPmeanX}(a) and (b) show the long time mean density profile for parameters for outside and just within the LR regime respectively. Also shown is the mean over the mesoscopic time $\tau$, i.e. a snapshot of the density profile from which the ``instantaneous'' DW position can be found. We note that in the LR regime (\fref{fig:SPmeanX}(b)) there are larger fluctuations in the $\tau$ averaged profile than for non-LR, and shape of the DW in the long time average profile is wider. The $\tau$ averaging method works well for small values of $\beta$; however for larger $\beta$ when the system is in the LR regime, determination of the position of the shock is much more difficult. This is because, due to the limited availability of the fuel carriers, the particles move more slowly -- the density fluctuations in the regions to the left and right of the DW exist on time scales similar to that of the movement of the wall. Any value of $\tau$ which will average out the microscopic fluctuations, will also average out the movement of the wall. This problem is compounded by the fact that the difference between the mean density on either side of the DW decreases with increasing $\beta$. Therefore we can only accurately measure the time course of the DW position just inside the LR regime; deep within that regime, we can only measure the mean wall position.

Figures \ref{fig:SPmeanX}(c)-(d), (e) and (f) show respectively typical time courses, normalised histograms of the wall positions, and the correlation function of the time course defined as
\begin{equation}
C(t')=\frac{\langle \delta x(t) \delta x(t') \rangle_t}{\langle \delta x(t)^2\rangle_t}, \label{tcor}
\end{equation}
where $\delta x(t)=x(t)-x$, and $\langle\cdots\rangle_t$ denotes average over time. The correlation functions are approximately exponential, and the examples shown have correlation time $2.3\times10^3~\mathrm{s}$ ($\beta=0.007~\mathrm{s}^{-1}$) and $3.6\times10^3~\mathrm{s}$ ($\beta=0.049~\mathrm{s}^{-1}$ - the LR regime). We find that whilst the time scale over which the DW moves is comparable in both cases, the width of the distribution is much wider in the LR case.

For the SP, previous studies \cite{Adams2007,Cook2010} have treated density fluctuations analytically by making the approximation that any fluctuations travel quickly along the lattice and are absorbed by the DW. That is to say, any fluctuation can be treated as a movement of the DW, and so only fluctuations in the rate at which particles move onto the $i=1$ and off of the $i=L$ sites need be considered. The fluctuations can then be described using a simple Langevin equation, leading to a power spectra $I\propto (\omega^2+\gamma^2)^{-1}$, where the constant $\gamma$ represents the restoring force which localises the DW.

\begin{figure}
\centering
\includegraphics{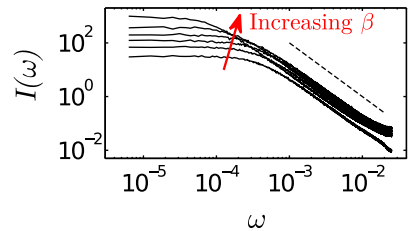}
\caption{Plot showing power spectra  of density fluctuations from simulations of systems in the SP with $\alpha_0=0.1~\mathrm{s}^{-1}$ and values of $\beta=0.01, ~0.02, ~0.03, ~0.04, ~0.05$ and $0.06~\mathrm{s}^{-1}$. The onset of the LR regime is at $\beta\sim0.04~\mathrm{s}^{-1}$, so the top 2 curves are for systems with LR. Other parameters are $L=500$, $V_0=20~\mathrm{s}^{-1}$ and $N=450$. The dashed line shows the slope $\omega^{-2}$.\label{fig:SPflucts} }
\end{figure}

\begin{figure}
\centering
\includegraphics{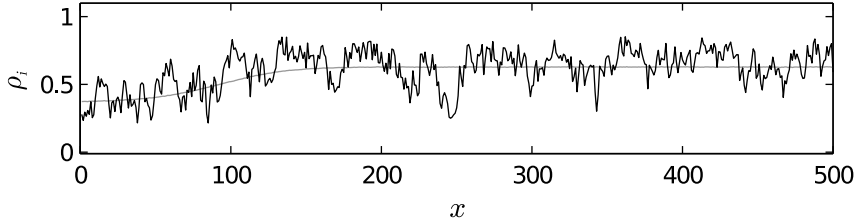}
\caption{Plot showing the short time average density profile for a system with LR in the SP (heavy line). The fluctuations are large and long lived enough that they are not averaged out; the DW is almost completely obscured. Also shown is the long time average profile (light line) where the DW is visible. Parameters are $\alpha_0=0.1125~\mathrm{s}^{-1}$, $\beta=0.063~\mathrm{s}^{-1}$, $V_0=20~\mathrm{s}^{-1}$, and $N=450$, and for the short time averaged profile $\tau=500~\mathrm{s}$. \label{fig:SPLRprofile} }
\end{figure}

\Fref{fig:SPflucts} shows that prediction of $I\propto \omega^{-2}$ for large $\omega$ still holds in the system with constrained fuel carriers. However, the value of $\gamma$ given by the theory does not correctly predict the behaviour at small $\omega$ for the LR regime. As noted above (and as can be seen by the increasing values of $I(\omega)$ with increasing $\beta$ in figure \ref{fig:SPflucts}) the fluctuations in the density increase significantly in the LR regime. These fluctuations are not quickly absorbed by movement of the front, but rather spend considerable time in other regions of the lattice; thus a theoretical treatment of fluctuations in a system with LR would require consideration of fluctuations in the hopping of particles at all sites, and is beyond the scope of the current work. The slow movement of the fluctuations along the lattice becomes particularly evident if we look at a snapshot of the density profile for a system deep within the LR regime (\fref{fig:SPLRprofile}); here the front seen in the time averaged profile is almost completely obscured by the fluctuations. 

In summary, we find that in the LR regime within the SP density fluctuations move more slowly through the lattice. This means that in this regime we cannot use the approximation that all fluctuations are absorbed by movements of the DW.

\section{Discussion and conclusion}

In this paper we have introduced and studied a TASEP model which has a constrained number of particles, as well as a constrained number of fuel carriers. That is to say, there is a finite rate of supply of the energy source which drives the system. 

In a system with only constrained particles \cite{Adams2008, Cook2009b, Cook2009,ciandrini_preparation_2010}, the coexistence or shock phase (SP) opens from a line at $\alpha=\beta$ to a region (i.e. a range of $\alpha_0$ and $\beta$ values). Also, for a very low number of particles ($N< LM/2$) the system cannot support an HD or MC phase. The introduction of a finite refuelling rate for the carriers leads to the existence of a limited resources (LR) regime within the LD, HD, MC and SP. As in a system with only constrained fuel carriers \cite{Brackley2010,Brackley2010b}, the LR regime is reached when the rate of fuel carrier use approaches that of the refuelling. 

The main characteristic of the extended model considering both finite particles and fuel carriers, is the existence of multiple phase transitions: through increasing only the parameter $\beta$, we obtain transitions from an HD phase to a shock phase, then \emph{back} to HD due to the onset of limited resources, then \emph{back again} to the SP before there is a transition to LD. This manifests as a cusp shape on the $\alpha_0$-$\beta$ phase plane. 

Finally we have analysed the fluctuations in the density, and have found them to be broadly in line with those seen in the unconstrained TASEP. For the range of parameters studied it appears that fluctuations in the number of loaded fuel carriers do not qualitatively change those in the particle density. We do find however that the speed at which fluctuations travel decreases in the LR regime, as would be expected due to the decrease in the mean hopping rate. This means that the oscillations seen in the power spectrum for the LD phase are damped out in the LR regime. The effect of a finite number of particles is the same as has been found in previous models \cite{Cook2009b,Cook2010}, i.e. the fluctuations are suppressed, particularly for low frequencies.

\ack

The authors would like to thank R J Allen, P Greulich, A Parmeggiani, M Thiel and I Stansfield for helpful discussions. Financial support was provided by BBSRC grants [BB/F00513/X1, BB/G010722] and the Scottish Universities Life Science Alliance (SULSA).

\providecommand{\newblock}{}

\end{document}